\documentclass[twocolumn,10pt,pra,aps,showpacs,longbibliography]{revtex4-1}
\usepackage{float}
\floatstyle{boxed}
\usepackage{wrapfig}
\usepackage{array}
\usepackage{graphicx}
\usepackage{amsbsy}
\usepackage{xcolor}
\usepackage[utf8x]{inputenc}
\usepackage{epstopdf}
\usepackage{amsmath,amssymb,amsfonts}

\begin{document}

\title{Complete identification of nonclassicality of Gaussian states via intensity moments}
\author{Ievgen I. Arkhipov}
\email{ievgen.arkhipov@gmail.com}
\affiliation{RCPTM, Joint Laboratory of Optics of Palack\'y University and
Institute of Physics of CAS, Faculty of Science, Palack\'y University, 17. listopadu
12, 771 46 Olomouc, Czech Republic}

\begin{abstract}
We present an experimental method for complete identification of  the nonclassicality of Gaussian states in the whole phase space. Our method  relies on nonclassicality witnesses written in terms of measured integrated intensity moments up to the third order, provided that  appropriate local coherent displacements are applied to the state under consideration.  The introduced approach, thus, only requires linear detectors for measuring intensities of optical fields, that is very convenient and powerful from the experimental point of view. Additionally, we demonstrate that the proposed technique not only allows to completely identify the nonclassicality of the Gaussian states but also to quantify it. 
\end{abstract}

\maketitle

Nonclassicality of light plays a crucial role in the field of quantum optics. The discovery of the nonclassical properties of light has led to the establishment of  new branches of quantum physics, e.g., to quantum information theory~\cite{NielsenBook}. 
One of the most known forms of the nonclassicality of light is the entanglement where different modes of quantum fields exhibit quantum correlations which have no analog in the classical optics~\cite{Einstein35,MandelBook,AgarwalBook}. The entangled states of light are now an indispensable source for quantum telecommunications and quantum computations~\cite{NielsenBook,ZeilingerBook}. 

The dimension of the Hilbert space of quantum systems can be either finite or infinite, as such, there are two most distinguished classes of the quantum states, namely, discrete variable  and continuous variable (CV) states, respectively. 
A lot of quantum protocols are now based on CV systems~\cite{Vaidman1994,Braunstein1998,Braunstein1998a,yonezawa04}. Moreover, Gaussian states, which are the subclass of CV states, possess the qualities of mathematical and experimental handiness, since the infinite Hilbert space for such states can be represented by a finite dimensional covariance matrix (CM). Additionally, the Gaussian states are easy to generate in a laboratory, e.g.,  they are the common output in the quantum parametric nonlinear processes~\cite{Perina1991Book}. 

The global nonclassicality of the Gaussian states of light can be expressed as  by the entanglement between optical modes and so by the local nonclassicalities in the form of field squeezing~\cite{Kimble77,Kimble1987,Paris97,Kim2002,Weedbrook12,Arkhipov2016b}.  It is worth noting that for discrete variable quantum systems one can also quantify the global nonclassicality in terms of local coherence and entanglement~\cite{Svozilik15,Cernoch2018}.

An important question thus arises, namely, how can one fairly certify the nonclassicality of the Gaussian states in the experiment? One of the solutions, although  experimentally complicated, can be a homodyne tomography~\cite{LeonhardtBook,Lvovsky2009}, which enables one to reconstruct the state in the phase space~\cite{Shchukin2006,Sperling2012a} and, thus, to identify the state's nonclassicality by knowing the form of the reconstructed   quasidistributing function. On the other hand, one would like to identify the nonclassicality without source and time-demanding reconstruction techniques. 
Apparently, a direct measurement of the intensity moments of optical fields can be utilized using easily accessable quadratric detectors or intensified CCD cameras which allow for  detecting nonclassicality of the weak and mesoscopic fields~\cite{Agarwal92,Mosset2004,Blanchet2008,Machulka2014}. Recently,  anomalous moments of the optical fields have been measured to identify the quantumness of light~\cite{Kuhn2017}. Also photon-number-resolving detectors yield a photocount histogram from which one can deduce the presence of the nonclassical correlations of the measured state with or without coherently displaced fields~\cite{Wallentowitz1996,Banaszek1999,Bondani2010,Pfister2014,Achilles2003,Fitch2003,Haderka2004,Avenhaus2010,Sperling2015,PerinaJr2017,PerinaJr2017a, Kiesel2012,Hadfield2009}. On the other hand by relying on the directly observed photocount statistics of the studied state one can lose the phase information of the system, since it may turn out that the nonclassicality of the state resides only  in the phase domain~\cite{Chan2007} or intensity measurements are simply unable to reveal the nonclassicality~\cite{Arkhipov2018a}. 
In parallel, by applying the dephased coherent displacement to the quantum field and using classical photodiodes one can retrieve intensity moments of higher orders from which the quantum properties of the given quantum state can be deduced, that is, the essence of an unbalanced homodying technique~\cite{Kuhn2016}.

In this paper, we show that by applying only local displacements to the Gaussian states  one can always identify the global nonclassicality of the state under consideration, i.e., its local squeezing and entanglement, by measuring only integrated intensity moments up to the second order.  Moreover, for a single- or two-mode squeezed light, the proposed method allows not only to identify its nonclassicality, but also even to quantify it. Our approach requires only  linear detectors since one has to measure coherently displaced fields, therefore one avoids the use of sophisticated detectors operating at the single-photon-level regime. Additionally, the presented method, as in the case of the unbalanced homodyne detection, is independent on  quantum efficiencies of the detectors, compared to the balanced homodyne technique. In short, we present a convenient experimental tool in the extraction of the nonclassical correlations of, in general,  the mixed  Gaussian states in the whole phase space. Furthermore, by utilizing the fact that the nonclassicality of the multimode Gaussian states can be expressed by means of single-mode auto- and two-mode cross-correlations and making use of the known multiport interferometric techniques~\cite{Rigovacca2016} we show that our approach  can be extended to the multimode case.


{\it Gaussian states and integrated intensity moments.}
Any $n$ -- mode  Gaussian state $\hat\rho$ can completely be characterized by its first and second statistical moments, i.e., by the average values of the operator  vector  $\boldsymbol{\hat X}=(\hat x_1,\hat p_1,\dots,\hat x_n,\hat p_n)^T$ and by the  covariance matrix $\boldsymbol{\hat\sigma}$ with elements $\sigma_{jk}=\frac{1}{2}\langle \hat X_j\hat X_k + \hat X_k\hat X_j\rangle-\langle\hat X_j\rangle\langle\hat X_k\rangle$, respectively, where the field quadratures $\hat x_l$ and $\hat p_l$ of the $l$th mode are related to the annihilation and creation operators $\hat a_l$ and $\hat a_l^{\dagger}$ as $\hat x_ l=1/\sqrt{2}(\hat a_l+\hat a_l^{\dagger})$, $\hat p_ l=-i/\sqrt{2}(\hat a_l-\hat a_l^{\dagger})$. 
The characteristic function,
\begin{equation} 
 \chi_{\hat\rho}(\boldsymbol{\Lambda})=\exp\left(-\frac{1}{2}\boldsymbol{\Lambda}^T\boldsymbol{\Omega}\boldsymbol{\sigma}\boldsymbol{\Omega}^T\boldsymbol{\Lambda}-i\boldsymbol{\Lambda}^T\boldsymbol{\Omega}\langle\boldsymbol{\hat X}\rangle\right),
\end{equation}
of  state $\hat\rho$  in phase space is of Gaussian form, where the vector  $\boldsymbol{\Lambda}=({\rm x_1},{\rm p_1},\dots,{\rm x_n},{\rm p_n})^T\in\mathbb{R}^{2n}$,  $\boldsymbol{\Omega}=\bigotimes\limits_{k=1}^{n}\omega_k$, and $\omega_k=\begin{pmatrix}0&1 \\ -1 & 0\end{pmatrix}$.  Moreover from the commutation relation $[\hat x_j,\hat p_k]=i\delta_{jk}$ it follows: 
\begin{equation}\label{HR}  
\boldsymbol{\sigma}+\frac{i}{2}\boldsymbol{\Omega}\geq0.
\end{equation}
The inequality in Eq.~(\ref{HR}) expresses the positivity of  state $\hat\rho$.

Introducing a new complex vector $\boldsymbol{\beta}=\boldsymbol{\Theta\Lambda}$, where $\boldsymbol{\Theta}=\bigotimes\limits_{k=1}^{n}\theta_k$, and $\theta_k=\frac{1}{\sqrt{2}}\begin{pmatrix}1&-i \\ 1 & i\end{pmatrix}$, one can arrive at the normal characteristic function $C_{\cal N}(\boldsymbol{\beta})= \chi_{\hat\rho}(\boldsymbol{\Theta}^{-1}\boldsymbol{\beta})\exp(\frac{1}{2}\boldsymbol{\beta^{\dagger}\beta})$, or explicitly,
\begin{equation}\label{CN} 
C_{\cal N}(\boldsymbol{\beta})=\exp\left(-\frac{1}{2}\boldsymbol{\beta}^{\dagger}\boldsymbol{\Omega}\boldsymbol{ A}_{\cal N}\boldsymbol{\Omega}^T\boldsymbol{\beta}+\boldsymbol{\beta}^{\dagger}\boldsymbol{\Omega}\boldsymbol{\Xi}\right),
\end{equation}
where $[\boldsymbol{A}_{\cal N}]_{jk}=\langle :\!\!\Delta\hat A_j^{\dagger} \Delta\hat A_k\!\!:\rangle =\langle :\!\! \hat A_j^{\dagger} \hat A_k\!\!:\rangle - \langle \hat A_j^{\dagger}\rangle\langle \hat A_k\rangle$ are the elements of the normal covariance matrix, and $\hat A = (\hat a_1^{\dagger},\hat a_1,\dots,\hat a_n^{\dagger},\hat a_n)^T$ is a vector of boson operators. The symbol $:\ :$ accounts for normal ordering of operators, i.e., all creation operators $\hat a^{\dagger}$ are put to the left with respect to annihilation operators $\hat a$. The complex vector $\boldsymbol{\Xi}=(\xi_1,\xi_1^*,\dots,\xi_n,\xi_n^*)^T\in{\mathbb C}^{2n}$ is in general a vector of displaced coherent fields. 

The normal generating function for the $n$-mode Gaussian state is given as
\begin{eqnarray}\label{Gdef} 
\hspace{-4mm}G_{\cal N}(\boldsymbol{\lambda}) =&&\frac{1}{\pi^n}\int C_{\cal N}(\boldsymbol{\beta})\prod\limits_{j=1}^n(\lambda_j)^{-1} \exp\left(-\frac{|\beta_j|^2}{\lambda_j}\right){\rm d}^2\beta_j, 
\end{eqnarray}
where $\boldsymbol{\lambda}=(\lambda_1,\dots,\lambda_n)\in{\mathbb R}^{n}$ is a real vector.

Combining now Eqs.~(\ref{CN}) and (\ref{Gdef}) we acquire
\begin{eqnarray}  
G_{\cal N}(\boldsymbol{\lambda})=\frac{1}{\sqrt{{\rm det}\boldsymbol{ A'_{\cal N}}}\prod\limits_{j=1}^n\lambda_j}\exp\left(-\frac{1}{2}\boldsymbol{\Xi^{\dagger}}\boldsymbol{{ A}'_{\cal N}}^{-1}\boldsymbol{\Xi}\right),
\end{eqnarray}
with $\boldsymbol{ A'_{\cal N}}=\boldsymbol{ A_{\cal N}}+\boldsymbol{\lambda}^{-1}\mathbb{I}_{2n}$, where $\mathbb{I}_{2n}$ is an identity matrix of dimension $2n$, and we denote the matrix $\boldsymbol{\lambda}^{-1}={\rm diag}(1/\lambda_1,1/\lambda_1,\dots,1/\lambda_n,1/\lambda_n)$.

The integrated intensity moments $ \langle W_1^{k_1}\dots W_n^{k_n}\rangle
$ are obtained along the formula,
\begin{eqnarray} 
\hspace{-3mm} \langle W_1^{k_1}\dots W_n^{k_n}\rangle &&= (-1)^{k_1+\dots+k_n} \nonumber \\
&&\times\left.\frac{\partial^{k_1+\dots+k_n}  G_{\cal   N}(\boldsymbol{\lambda})}{\partial\lambda_1^{k_1}\dots\partial\lambda_n^{k_n}}\right|_{\lambda_1=\dots=\lambda_n=0}.
\end{eqnarray}


{\it Nonclassicality criteria based on integrated intensity moments.}
The very form of the covariance matrix of the $n$--mode Gaussian state suggests that all the nonclassicality properties are encoded into single-mode squeezing of each mode and entanglement between two arbitrary modes of the state. Indeed, looking at the normal CM,
\begin{eqnarray}\label{multiCM} 
\boldsymbol{A_{\cal N}}=\begin{pmatrix}
{\bf A_1} & {\bf A_{12}} & \cdots & {\bf A_{1n}} \\
{\bf A_{12}^{\dagger}} & {\bf A_{2}} & \cdots & \vdots \\
\vdots & \vdots & \ddots & \vdots \\
{\bf A_{1n}^\dagger} & \cdots & \cdots & {\bf A_n}
\end{pmatrix}.
\end{eqnarray}
where $\bf A_k$ and $\bf A_{jl}$ are block $2\times2$ matrices 
\begin{eqnarray}\label{lcoef}  
{\bf A_k}=\begin{pmatrix}
B_k & C_k \\
C_k^* & B_k
\end{pmatrix}, \quad 
\begin{matrix} 
B_k&=&\langle:\!\Delta\hat a_k^{\dagger}\Delta\hat a_k\!:\rangle, \\
C_k &=& \langle:\!\Delta\hat a_k^{2}\!:\rangle,
\end{matrix}
\end{eqnarray}
\begin{eqnarray}\label{icoef}  
{\bf A_{jl}}=\begin{pmatrix}
\bar D^*_{jl} & D_{jl} \\
D^*_{jl} & \bar D_{jl}
\end{pmatrix}, \quad 
\begin{matrix}
D_{jl}&=&\langle:\!\Delta\hat a_j\Delta\hat a_l\!:\rangle, \\
\bar D_{jl}&=&\langle:\!\Delta\hat a_j^{\dagger}\Delta\hat a_l\!:\rangle, \\
\end{matrix}
\end{eqnarray}
describing the quantum auto-correlation of mode $k$ and cross-correlations between modes $j$ and $l$, respectively,
one can see that the single- and two-mode correlations explicitly determine the $n$-mode Gaussian state. Indeed, by means of appropriate unitary operations one can always reduce the $n$-mode nonclassical Gaussian state to the separable $n$-mode locally squeezed states or, as in the case of the pure Gaussian states, to the tensor product of $\frac{n}{2}$  ($n=2k, k\in {\mathbb Z}$) two-mode squeezed states~\cite{Braunstein05,Eisert2003}. Below we show that such single-mode and two-mode  nonclassicality correlations of multimode Gaussian states can be retrieved by integrated intensity moments up to the second order.

The single-mode nonclassicality witness (NW)  expressed in terms of integrated intensity moments can be written as the following~\cite{Lee1990a}:
\begin{eqnarray}\label{R} 
R_k= \langle W_k\rangle\langle W_k^3\rangle - \langle W_k^2\rangle^2<0, \quad k = 1,\dots,n.
\end{eqnarray}
Whenever $R_k<0$  a mode $k$ exhibits nonclassicality in the form of squeezing.

To quantify the nonclassicality of two modes the following NW can be used~\cite{PerinaJr2017a}:
\begin{eqnarray}\label{M}  
M_{jl} = \langle W_j^{2}\rangle \langle W_l^2\rangle - \langle W_jW_l\rangle^2<0.
\end{eqnarray}
Whenever $M_{jl}<0$ the two-mode Gaussian state cannot be both locally classical and separable.

It is important to note that the NWs in Eqs.~(\ref{R}) and (\ref{M}) can be used for the detection of nonclassicality for any kind of state of light, i.e., even for non-Gaussian states since the negativity of these NWs refers to nonclassical properties of the quasidistribution Glauber-Sudarshan $P$ function~\cite{AgarwalBook,Shchukin2006,PerinaJr2017a}. But they become optimal for complete nonclassicality detection only for Gaussian states.

Now, we would like to show that NWs $R_k$ and $M_{jl}$ can be used as genuine nonclassicality identifiers for single- and two-mode Gaussian states, provided that  appropriate local unitary operations are applied to the corresponding modes. Moreover, we will demonstrate that the NW $M_{jl}$ is useful even for verification of the nonclassicality of the single-mode states.


{\it Theorem 1.} {A local nonclassicality of the single-mode Gaussian state $\rho_k$ can be revealed and quantified by the nonclassicality witness $R_k$ by means of the appropriate coherent displacement operator of the given mode.}

{\it Proof.} By applying a coherent displacement operator $\hat D(\xi'_k)=e^{\hat a_k^{\dagger}\xi'_k-\hat a_k\xi'^*_k}$ to the single-mode Gaussian state $\hat\rho_k$, i.e.,  $\hat\rho_k\to\hat\rho'_k=\hat D(\xi'_k)\hat\rho_k\hat D(\xi'_k)^{\dagger}$ such that the coherent shifting vector $\boldsymbol\Xi$ in the normal characteristic function $C_{\cal N}[\hat\rho'_k]$ takes the form $\boldsymbol{\Xi}\to\boldsymbol{\Xi'}=(|\xi|e^{i\alpha_k},|\xi|e^{-i\alpha_k})$,  one then can rewrite the NW $R_k$ for the state $\hat\rho'_k$ in the polynomial form
\begin{equation}\label{pol}  
R_k=ax^3+bx^2+cx+d, 
\end{equation}
 where $x=|\xi|^2$, and $a,b,c,d,$ are functions of  $B_k$,  $C_k$, which are  given in Eq.~(\ref{lcoef}) and of  phase $\alpha_k$. The condition at which $R_k$  may acquire negative values can automatically be satisfied whenever $a= 2(B_k+{\rm Re}[C_ke^{-2i\alpha_k}])<0$ since in that case one can always find such  $x\in[0,\infty)$ for which $R_k<0$. The complex parameter $C_k$ can be presented as $C_k=|C_k|e^{i\phi_k}$, and by setting $\alpha_k=1/2(\phi_k-\pi)$ one arrives at 
\begin{equation}\label{snon}  
a \equiv B_k-|C_k|. 
\end{equation}
The expression in Eq.~(\ref{snon}) is nothing else but the condition of the Gaussian single-mode nonclassicality when negative~\cite{Arkhipov2016b, Arkhipov2016c}. Moreover, the negative values of $a$ in Eq.~(\ref{snon}) are a monotone of  Lee's nonclassicality depth $\tau$, which is a good nonclassicality monotone for the Gaussian states~\cite{Arkhipov2016b,Lee1991}. Therefore, if  state   $\hat\rho_k$ is nonclassical, one can always make NW $R_k$  not only detect its nonclassicality, but also quantify it. This completes the proof of the theorem.

When the free coefficient is 
\begin{equation*}
d=2B_k^4+5|C_k|^2B_k^2-|C_k|^4<0
\end{equation*}
 in Eq.~(\ref{pol}), then $R_k$ becomes negative even with $\boldsymbol\Xi'=0$. In that case the coherent displacement can be used for the enhancement of the nonclassicality detection since $\lim\limits_{|\xi_k|\rightarrow\infty}R_k =-\infty$. 
If initially $d>0$ then by choosing such  $|\xi|>|\xi|_{cr}$, one eventually can access the negative values of $R_k$.  The critical values $|\xi|_{cr}$ are found as one of the real positive roots of the NW $R_k$,
\begin{equation}\label{cr}  
|\xi|_{cr}=\left[-1/3a(b+\eta^hF+\Delta_0/\eta^hF)\right]^{1/2}, \quad h=0,1,2,
\end{equation}
where 
\begin{equation}
\eta=\frac{-1+i\sqrt{3}}{2}, \quad  F=\sqrt[3]{\frac{\Delta_1\pm\sqrt{\Delta_1^2-4\Delta_0^2}}{2}},  \nonumber 
\end{equation}
and $\Delta_0=b^2-3ac$,  $\Delta_1=2b^3-9abc+27a^2d$.

{\it Lemma 1.} The NW $M_{jl}$ in Eq.~(\ref{M}) is invariant with respect to the local phase shifting operations $\hat S=\hat S_j(\phi_j)\otimes\hat S_l(\phi_l)$ applied to  modes $j$ and $l$ of the two-mode Gaussian state $\hat\rho_{jl}$.

{\it Proof.} The local phase-shifting operations $\hat S=\hat S_j(\phi_j)\otimes\hat S_l(\phi_l)$ applied to the two-mode Gaussian state $\hat\rho_{jl}$ transform the corresponding  boson operators of the modes as $\hat a_k\rightarrow\hat a_ke^{i\phi_k}$, where $k=j,l$. By putting the latter into  Eq.~(\ref{M}) and utilizing the following expression for integrated intensity moments $\langle W_j^mW_l^n\rangle=\langle \hat a_j^{\dagger m}\hat a_l^{\dagger n}\hat a_j^m\hat a_l^n\rangle$, one makes sure that NW $M_{jl}$ remains unchanged under such transformations.

{\it Theorem 2.} 
The nonclassicality of the two-mode Gaussian state $\hat\rho_{jl}$ expressed solely in the form of the entanglement  can be completely detected by the nonclassicality witness  $M_{jl}$ provided that appropriate  local coherent displacements  are applied to the state.

{\it Proof.}
With the appropriate combination of the local phase shifting $\hat S=\hat S_j\otimes\hat S_l$ and coherent displacements $\hat D=\hat D_j(\xi'_j)\hat D_l(\xi'_l)$, where subscripts $j$ ($l$) denote an operator acting on $j$th ($l$th) mode of the entangled two-mode Gaussian state $\hat\rho_{jl}$, one can transform  state $\hat\rho_{jl}\to\hat\rho'_{jl}\equiv\hat D\hat S\hat\rho_{jl}\hat S^{\dagger}\hat D^{\dagger}$  such, that  the normal covariance matrix $\boldsymbol{ A'_{\cal N}}=\hat S_{jl}\boldsymbol{ A_{\cal N}}\hat S_{jl}^{\dagger}$ attains real nonzero elements $B_j$, $B_l$, $\bar D_{jl}$, and $D_{jl}$ with $C_j=C_l=0$ by default (for nonzero $C_j$ and $C_l$ the proof is straightforward), and the coherent shifting vector $\boldsymbol{\Xi}\to\boldsymbol{\Xi'}$  in the normal characteristic function $C_{\cal N}[\hat\rho'_{jl}]$  becomes of the form  $\boldsymbol{\Xi'}=(|\xi|e^{i\alpha_j},|\xi|e^{-i\alpha_j},|\xi|e^{i\alpha_l},|\xi|e^{-i\alpha_l})$, i.e., with equal amplitudes but different phases. The form of the matrix $\boldsymbol{ A'_{\cal N}}$ is such that it corresponds to the standard form of the symmetrical covariance matrix  $\boldsymbol{\sigma_{\rm st}}$,
\begin{eqnarray}\label{sigmast} 
\boldsymbol{\sigma_{\rm st}}=\begin{pmatrix}
q_j & 0 & q_{jl} & 0 \\
0 & q_j & 0 & q'_{jl} \\
q_{jl} & 0 & q_l & 0 \\
0 & q'_{jl} & 0 & q_l
\end{pmatrix},
\end{eqnarray}
i.e., the applied phase shifting operation $\hat S$ does not affect the global nonclassicality of the state~\cite{Duan2000}.
The relations between the elements of $\boldsymbol{\sigma_{\rm st}}$ and $\boldsymbol{ A'_{\cal N}}$ are given as the following:
\begin{eqnarray}\label{stf}  
B_k &=& q_k-\frac{1}{2}, \quad k = j,l \nonumber \\
D_{jl} &=& \frac{q_{jl}-q'_{jl}}{2}, \quad \bar D_{jl} = \frac{q_{jl}+q'_{jl}}{2}.
\end{eqnarray}
\noindent
 The NW $M_{jl}$ in Eq.~(\ref{M}) for state $\hat\rho'_{jl}$ with the normal characteristic function $C_{\cal N}[\hat\rho'_{jl}]$  then can be presented as a polynomial $M_{jl}=M_{jl}(x)$ with $x=|\xi|^2$, which has the same form as in Eq.~(\ref{pol}),  where $a,b,c,d,$ now are real functions of $\alpha_j$, $\alpha_l$ and $B_j$, $B_l$, $\bar D_{jl}$, and $D_{jl}$.
A sufficient condition for which $M_{jl}(x)$ can acquire negative values for nonclassical state $\hat\rho_{jl}$  is  when the coefficient $a$ is  negative, i.e.,
\begin{eqnarray}\label{Ma}  
a\equiv B_j+B_l-2D_{jl}\cos(\alpha_j+\alpha_l)+2\bar D_{jl}\cos(\alpha_j-\alpha_l)<0. \nonumber \\
\end{eqnarray}
If one puts $\alpha_q=(2k_q+1)\pi/4$, $k_q\in{\mathbb Z}$, $q=j,l$, into Eq.~(\ref{Ma}) the coefficient $a$ becomes equivalent to the condition for inseparability of the state. Indeed, in that case  Eq.~(\ref{Ma}) along with  Eq.~(\ref{stf}) can be rewritten as 
\begin{equation}\label{nsepduan}  
\langle(\Delta\hat u)^2\rangle+\langle(\Delta\hat v)^2\rangle<1,
\end{equation}
 where $\hat u = (|h|\hat x_j+\hat x_l/h)/\sqrt{2}$, $\hat v = (|h|\hat p_j-\hat p_l/h)/\sqrt{2}$, and $h=\pm 1$.  Equation~(\ref{nsepduan}) is the inseparability condition written for fields' quadratures and was first derived in Ref.~\cite{Duan2000}. Now, by applying the inverse phase-shifting operations $\hat S^{-1}$ to the modified state $\hat\rho_{jl}'$ in order to come back to the initial form of the covariance matrix $\boldsymbol{A_{\cal N}}$  and by making use of  {\it Lemma 1}, one concludes that for any entangled two-mode Gaussian state $\hat\rho_{jl}$  there are always such local coherent displacements for which the NW $M_{jl}$ is able to detect the entanglement of the state. This completes the proof of the theorem.


If some coherent displacement operations are applied to the state, i.e., $\hat\rho'_{jl}=\hat D\hat\rho_{jl}\hat D^{\dagger}$, meaning that only the vector $\boldsymbol{\Xi}\to\boldsymbol{\Xi'}$  is modified in the normal characteristic function $C_{\cal N}[\hat\rho'_{jl}]$, and as before $\boldsymbol{\Xi'}=(|\xi|e^{i\alpha_j},|\xi|e^{-i\alpha_j},|\xi|e^{i\alpha_l},|\xi|e^{-i\alpha_l})$, then the coefficient $a$ in $M_{jl}(x=|\xi|^2)$ takes the following form:
\begin{eqnarray}\label{aM}  
a\equiv&& B_j+B_l+{\rm Re}[C_je^{-2i\alpha_j}]+{\rm Re}[C_le^{-2i\alpha_l}] \nonumber \\
&&-2{\rm Re}[\bar D_{jl}e^{i(\alpha_j-\alpha_l)}]-2{\rm Re}[D_{jl}e^{-i(\alpha_j+\alpha_l)}].
\end{eqnarray}
One can immediately see that in that case it is possible to retrieve the nonclassicality of the state by means of $M_{jl}$, which is expressed by the entanglement if choosing the right phases of the displaced coherent fields. 

For instance, for two-mode squeezed vacuum states or so-called twin beam states which exhibit entanglement, the Eq.~(\ref{aM}) can be reduced to 
\begin{equation}\label{nsep}  
a\equiv B_j+B_l-2|D_{jl}|,
\end{equation}
since for such states $C_j=C_l=\bar D_{jl}=0$, and where we define ${\rm arg}(D_{jl})=\alpha_j+\alpha_l$, ${\rm arg}$ stands for the argument of a complex number.
Equation~(\ref{nsep}) represents the entanglement condition for  twin beam states when negative~\cite{Arkhipov2015}. Additionally, the negative values of the coefficient $a$ in Eq.~(\ref{nsep}) are a monotone of the entanglement negativity, which is, by itself, a good entanglement monotone for two-mode Gaussian states~\cite{prepare}. Thus, for twin beams the NW $M_{jl}$ can also be used as an entanglement quantifier.

Most importantly, the NW $M_{jl}$ can detect, apart from the entanglement, also the local squeezing. For example, for the two-mode state $\hat\rho_{jl}$ which is both locally nonclassical and entangled then by suitably chosen $\alpha_j$ and $\alpha_l$ one can obtain the condition of the nonclassicality of such a state. 
 Namely, for $\alpha_j=\phi_j-\pi(n_2-n_1+1/2)$ and $\alpha_l=\phi_l+\pi(n_1+n_2+1/2)$, where $\phi_j$ ($\phi_l$) is the phase of the boson operator $\hat a_j$ ($\hat a_l$) of the mode $j$ ($l$) and $n_1,n_2\in{\mathbb Z}$, Eq.~(\ref{aM}) attains the form
\begin{equation}\label{noncom}  
a\equiv B_j-|C_j|+B_l-|C_l|-2(|D_{jl}|-|\bar D_{jl}|).
\end{equation}
Equation~(\ref{noncom}) is especially easy to read for pure Gaussian states since the first four terms in  Eq.~(\ref{noncom}) are responsible as before for the local nonclassicality, the differences in the last two terms in the parentheses are responsible for the entanglement~\cite{Arkhipov2016b,Arkhipov2016c}, and thus $a<0$ for such states. 

Moreover, the form of the coefficient $a$ in Eq.~(\ref{aM}) suggests that the NW $M_{jl}$ allows for detecting the nonclassicality of the single-mode state. For example, to detect the nonclassicality of the single-mode $j$ one needs to replace mode $l$ just with the coherent field $\xi_l=|\xi|e^{i\alpha_l}$, where $\alpha_l$ can be arbitrary since in that case  Eq.~(\ref{aM}) reduces to Eq.~(\ref{snon}).

We would also like to note that the coherent displacements of the Gaussian states generated in the spontaneous parametric processes could be encompassed by means of the stimulated emission of the corresponding parametric process. As such, by means of an appropriate choice of the initial phase and intensity of the stimulating coherent fields one can completely reveal the nonclassicality of the given Gaussian state~\cite{arkhipov2018d}.

The author thanks J. Pe\v{r}ina Jr. and A. Miranowicz for valuable discussions.  This research was supported by  GA \v{C}R Project  No.~17-23005Y.


%

\end{document}